\documentclass[aps,prl,twocolumn,showpacs,floatfix,nofootinbib]{revtex4}
\usepackage{amsmath,graphicx}
\newcommand{\mean}[1]{\left\langle #1 \right\rangle}
\begin{document}

\title{Complete relativistic second-order dissipative hydrodynamics from the entropy principle}

\author{Amaresh Jaiswal, Rajeev S. Bhalerao and Subrata Pal}
\affiliation{Tata Institute of Fundamental Research,
Homi Bhabha Road, Mumbai 400005, India}

\date{\today}

\begin{abstract}

We present a new derivation of relativistic dissipative hydrodynamic
equations, which invokes the second law of thermodynamics for the
entropy four-current expressed in terms of the single-particle
phase-space distribution function obtained from Grad's 14-moment
approximation. This derivation is complete in the sense that all the
second-order transport coefficients are uniquely determined within a
single theoretical framework. In particular, this removes the
long-standing ambiguity in the relaxation time for bulk viscosity
thereby eliminating one of the uncertainties in the extraction of the
shear viscosity to entropy density ratio from confrontation with the
anisotropic flow data in relativistic heavy-ion collisions. We find
that in the one-dimensional scaling expansion, these transport
coefficients prevent the occurrence of cavitation even for rather
large values of the bulk viscosity estimated in lattice QCD.

\end{abstract}

\pacs{25.75.Ld, 24.10.Nz, 47.75+f}


\maketitle

Relativistic fluid dynamics has been quite successful in explaining
the various collective phenomena observed in cosmology, astrophysics
and the physics of high-energy heavy-ion collisions. The earliest
theories of relativistic dissipative hydrodynamics by Eckart
\cite{Eckart:1940zz} and Landau-Lifshitz \cite{Landau} were based on
the assumption that the entropy four-current is first order in
dissipative quantities, which led to parabolic differential equations
that suffered from acausality. The second-order Israel-Stewart (IS)
theory \cite{Israel:1979wp} with the entropy current quadratic in
dissipative quantities led to hyperbolic equations and thus restored
causality.

Application of the second-order viscous hydrodynamics to high-energy
heavy-ion collisions has evoked widespread interest ever since a
surprisingly small value for the shear viscosity to entropy density
ratio $\eta/s$ was estimated from the analysis of the elliptic flow
data \cite{Romatschke:2007mq}. Indeed the estimated $\eta/s$
was close to the conjectured lower bound $\eta/s|_{\rm KSS} = 1/4\pi$
\cite{Policastro:2001yc,Kovtun:2004de}. This led to the claim that the
quark-gluon plasma (QGP) formed at the Relativistic Heavy-Ion Collider
(RHIC) was the most perfect fluid ever observed. A precise estimate of
$\eta/s$ is vital to the understanding of the properties of the
QCD matter.

In this Communication, we provide a solution to one of the major
uncertainties that hinders an accurate extraction of the viscous
corrections to the ideal fluid behavior, namely the inadequate
knowledge of the second-order transport coefficients. In the standard
derivation of second-order evolution equations for dissipative
quantities from the requirement of positive divergence of the entropy
four-current, the most general algebraic form of the entropy current
is parameterized in terms of unknown thermodynamic coefficients
\cite{Israel:1979wp}. These coefficients which are related to
relaxation times and coupling lengths of the shear and bulk pressures
and heat current, however, remain undetermined within the framework of
thermodynamics alone \cite{Muronga:2003ta}. While kinetic theory for
massless particles \cite{Baier:2006um} and strongly coupled 
${\cal N}=4$ supersymmetric Yang-Mills theory \cite{Baier:2007ix}
predict different shear relaxation times $\tau_\pi = 3/2\pi T$ and
$(2-\ln 2)/2\pi T$, respectively, for $\eta/s = 1/4\pi$, the bulk
relaxation time $\tau_\Pi$ remains completely ambiguous. Hence ad hoc
choices have been made for the value of $\tau_\Pi$ in hydrodynamic
studies
\cite{Fries:2008ts,Denicol:2009am,Song:2009rh,Rajagopal:2009yw}.

Lattice QCD studies for gluonic plasma in fact predict large values of
bulk viscosity to entropy density ratio, $\zeta/s$, of about (6-25)
$\eta/s|_{\rm KSS}$ near the QCD phase-transition temperature $T_c$
\cite{Meyer:2007dy}. This would translate into large values of the
bulk pressure and bulk relaxation time, and may affect the evolution
of the system significantly
\cite{Denicol:2009am,Song:2009rh}. Further, the large bulk pressure
could result in a negative longitudinal pressure leading to mechanical
instabilities (cavitation) whereby the fluid breaks up into droplets
\cite{Torrieri:2008ip,Rajagopal:2009yw,Bhatt:2010cy}. Thus the theoretical
uncertainties arising from the absence of reliable estimates for the
second-order transport coefficients should be eliminated for a proper
understanding of the system evolution.

We present here a formal derivation of the dissipative
hydrodynamic equations where all the second-order transport
coefficients get determined uniquely within a single theoretical
framework. This is achieved by invoking the second law of
thermodynamics for the generalized entropy four-current expressed in
terms of the phase-space distribution function given by Grad's
14-moment approximation. Significance of these coefficients is
demonstrated in one-dimensional scaling expansion of the viscous
medium.

Hydrodynamic evolution of a medium is governed by the conservation
equations for the energy-momentum tensor and particle current
\cite{deGroot}
\begin{align}\label{NTD}
T^{\mu\nu} &= \int dp \ p^\mu p^\nu (f+\bar f) = \epsilon u^\mu u^\nu-(P+\Pi)\Delta ^{\mu \nu} 
+ \pi^{\mu\nu},  \nonumber\\
N^\mu &= \int dp \ p^\mu (f-\bar f) = nu^\mu + n^\mu,
\end{align}
where $dp = g d{\bf p}/[(2 \pi)^3\sqrt{{\bf p}^2+m^2}]$, $g$ and $m$
being the degeneracy factor and particle rest mass, $p^{\mu}$ is the
particle four-momentum, $f\equiv f(x,p)$ is the phase-space
distribution function for particles and $\bar f$ for antiparticles.
The above integral expressions assume the system to be dilute so that 
the effects of interaction are small \cite{deGroot}.
In the above tensor decompositions, $\epsilon,
P, n$ are respectively energy density, pressure, net number density, and
the dissipative quantities are the bulk viscous pressure $(\Pi)$,
shear stress tensor $(\pi^{\mu\nu})$ and particle diffusion current
$(n^\mu)$. Here $\Delta^{\mu\nu}=g^{\mu\nu}-u^\mu u^\nu$ is the
projection operator on the three-space orthogonal to the hydrodynamic
four-velocity $u^\mu$ defined in the Landau frame: $T^{\mu\nu}
u_\nu=\epsilon u^\mu$. 

Energy-momentum conservation, $\partial_\mu T^{\mu\nu} =0$ and current
conservation, $\partial_\mu N^{\mu}=0$ yield the fundamental evolution
equations for $\epsilon$, $u^\mu$ and $n$.
\begin{align}\label{evol}
D\epsilon + (\epsilon+P+\Pi)\partial_\mu u^\mu - \pi^{\mu\nu}\nabla_{(\mu} u_{\nu)} &= 0,  \nonumber\\
(\epsilon+P+\Pi)D u^\alpha - \nabla^\alpha (P+\Pi) + \Delta^\alpha_\nu \partial_\mu \pi^{\mu\nu}  &= 0,  \nonumber\\
Dn + n\partial_\mu u^\mu + \partial_\mu n^{\mu} &=0.
\end{align}
We use the standard notation $A^{(\alpha}B^{\beta )} = (A^\alpha
B^\beta + A^\beta B^\alpha)/2$, $D=u^\mu\partial_\mu$, and
$\nabla^\alpha = \Delta^{\mu\alpha}\partial_\mu$. Even if the
equation of state is given, the system of Eqs. (\ref{evol}) is not
closed unless the evolution equations for the dissipative quantities
$\Pi$, $\pi^{\mu\nu}$, $n^\mu$ are specified.

Traditionally the dissipative equations have been obtained by invoking
the second law of thermodynamics, viz., $\partial_\mu S^\mu \geq 0$,
where the entropy four-current $S^\mu$ is given by 
\cite{Israel:1979wp,Baier:2006um,Muronga:2003ta}
\begin{align}\label{AEFC}
S^\mu =& \ P\beta u^\mu - \alpha N^\mu + \beta u_\nu T^{\mu \nu}-Q^\mu(\delta N^\mu, \delta T^{\mu \nu}) \nonumber \\
=& \ s u^\mu -
\frac{\mu n^\mu}{T}-
 \left(\beta_0\Pi^2 - \beta_1 n_\nu n^\nu 
+ \beta_2\pi_{\rho\sigma} \pi^{\rho\sigma}\right) \frac{u^\mu}{2T} \nonumber\\
&- \left(\alpha_0\Pi\Delta^{\mu\nu} + \alpha_1\pi^{\mu\nu}\right)\frac{n_\nu}{T}.
\end{align}
Here $\beta=1/T$ is the inverse temperature, $\mu$ is the chemical
potential, $\alpha=\beta \mu$, and $Q^\mu$ is a function of deviations
from local equilibrium. The second equality is obtained by
using the definition of the equilibrium entropy density
$s=\beta (\epsilon+P-\mu n)$ and 
Taylor-expanding $Q^\mu$ to second order in dissipative fluxes.
In this expansion,
$\beta_i(\epsilon,n) \geq 0$ and $\alpha_i(\epsilon,n) \geq 0$ are the
thermodynamic coefficients corresponding to pure and mixed terms. 
These coefficients can be obtained within the kinetic theory approach
such as the IS theory \cite{Israel:1979wp}. However, it
is important to note that they cannot be determined
solely from thermodynamics using Eq. (\ref{AEFC}) and as a consequence 
the evolution equations remain incomplete.

In contrast to the above approach, our starting point for the
derivation of the dissipative evolution equations is the entropy
four-current expression generalized from Boltzmann's H-function:
\begin{eqnarray}\label{EFC}
S^\mu_{r=0} &=& -\int dp ~p^\mu \left[ f \left(\ln f - 1\right)+(f \to \bar f) \right] , \nonumber \\  
S^\mu_{r=\pm 1} &=& -\int dp ~p^\mu \left[ \left(f \ln f + r\tilde f \ln \tilde f\right)+(f \to \bar f) \right],\quad
\end{eqnarray}
where $\tilde f \equiv 1 - rf$ and $r = 1,-1,0$ for Fermi, Bose, and
Boltzmann gas, respectively. The divergence of $S^\mu_{r=0,\pm 1}$ leads to
\begin{equation}\label{EFCD}
\partial_\mu S^\mu = -\int dp ~p^\mu \left[ \left(\partial_\mu f\right) 
\ln (f/\tilde f) +(f \to \bar f) \right] .
\end{equation}

For small departures from equilibrium, $f$ and $\bar f$ 
can be written as $f =
f_0 + \delta f$ and $\bar f = \bar f_0 + \delta \bar f$.
The equilibrium distribution functions are defined as
$f_0 = [\exp(\beta u\cdot p -\alpha) + r]^{-1}$ and
$\bar f_0 = [\exp(\beta u\cdot p +\alpha) + r]^{-1}$, where $\beta=1/T$ and
$\alpha=\mu/T$ are obtained from the equilibrium matching conditions
$n\equiv n_0$ and $\epsilon \equiv \epsilon_0$.

To proceed further, we take recourse to Grad's 14-moment approximation
\cite{Grad} for the single particle distribution in orthogonal basis
\cite{Denicol:2010xn,Jaiswal:2012qm}
\begin{equation}\label{G14}
f = f_0 + f_0 \tilde f_0 \phi, ~~~ \phi =  \lambda_\Pi \Pi + \lambda_n n_\alpha p^\alpha 
+ \lambda_\pi \pi_{\alpha\beta} p^\alpha p^\beta, \quad
\end{equation}
and similarly for $\bar f$.
The coefficients ($\lambda_\Pi, \lambda_n, \lambda_\pi$) are assumed 
to be independent of four-momentum $p^{\mu}$ and are functions of 
$(\epsilon, \alpha, \beta)$. From Eqs. (\ref{EFCD}) and (\ref{G14}), we get
\begin{align}\label{EFCD1}
\partial_\mu S^\mu = -\!\!\int\! dp ~ p^\mu \Big[
& \left(\partial_\mu f\right)
\left\{ \ln \! \left( \frac{f_0}{\tilde f_0} \right)\! + \ln \!
\left(\! 1\! + \frac{\phi}{1-rf_0\phi}\right) \! \right\} \nonumber \\
& + (f \to \bar f, ~ f_0 \to \bar f_0) \Big].
\end{align}
The $\phi$-independent terms on the right vanish due to 
energy-momentum and current
conservation equations. To obtain second-order evolution equations for
dissipative quantities, one should consider $S^\mu$ up to the same
order. Hence $\partial_\mu S^\mu$ necessarily becomes third-order.
Expanding the $\phi$-dependent terms
in Eq. (\ref{EFCD1}) and retaining all terms
up to third order in gradients (where $\phi$ is linear in dissipative 
quantities), we get
\begin{align}\label{EFCD2}
\partial_\mu S^\mu = & -\int  dp ~p^\mu \Big[ \Big \{
\phi\left(\partial_\mu f_0\right) - 
\phi^2 (\tilde f_0 -1/2)(\partial_{\mu}f_0) \nonumber \\
&+\phi^2 \partial_\mu (f_0 \tilde f_0) + \phi f_0 \tilde f_0 (\partial_\mu \phi) \Big \}
+ (f_0 \to \bar f_0) \Big].
\end{align}
The various integrals in the above equation can be decomposed into 
hydrodynamic tensor degrees of freedom via the definitions:
\begin{align}\label{TDI}
I^{\mu_1\mu_2\cdots\mu_n}_\pm \equiv& \int dp \ p^{\mu_1}  \cdots p^{\mu_n} (f_0 \pm \bar f_0) 
= I_{n0}^\pm u^{\mu_1}  \cdots u^{\mu_n} \nonumber \\
& + I_{n1}^\pm (\Delta^{\mu_1\mu_2} u^{\mu_3} \cdots u^{\mu_n} + \mathrm{perms}) + \cdots,
\end{align}
where `perms' denotes all non-trivial permutations of the Lorentz
indices. We similarly define $J^{\mu_1\mu_2\cdots\mu_n}_\pm $ and
$K^{\mu_1\mu_2\cdots\mu_n}_\pm $ where the momentum integrals are weighted
with $f_0 \tilde f_0 \pm (f_0 \to \bar f_0)$ and
$f_0 \tilde f_0^2 \pm (f_0 \to \bar f_0)$, 
and are tensor decomposed
with coefficients $J_{nq}^\pm$ and $K_{nq}^\pm$, respectively. 
All these coefficients can be
obtained by suitable contractions of the integrals and are related to
each other by
\begin{align}\label{ICR}
2K_{nq}^\pm &= J_{nq}^\pm + \frac{1}{\beta}\big[- J_{n-1,q-1}^\pm + (n-2q)J_{n-1,q}^\pm\big], \nonumber \\
J_{nq}^\pm &= \frac{1}{\beta}\left[-I_{n-1,q-1}^\pm + (n-2q)I_{n-1,q}^\pm \right],
\end{align}
and also satisfy the differential relations
\begin{align}\label{ICDR}
2K_{nq}^\pm &= J_{nq}^\pm - \frac{d}{d\beta}J_{n-1,q}^\pm = J_{nq}^\pm + \frac{d}{d\alpha}J_{nq}^\pm, \nonumber \\
J_{nq}^\pm &= -\frac{d}{d\beta}I_{n-1,q}^\pm = \frac{d}{d\alpha}I_{nq}^\pm. 
\end{align}
With the help of these relations and Grad's 14-moment approximation,
Eq. (\ref{EFCD2}) reduces to
\begin{align}\label{EFCD3}
\partial_\mu S^\mu = 
& -\! \beta\Pi\Big[ \theta +\! \beta_0\dot\Pi 
+\! \beta_{\Pi\Pi} \Pi\theta 
+ \alpha_0 \nabla_\mu n^{\mu} 
+ \psi\alpha_{n\Pi } n_\mu \dot u^\mu \nonumber \\
&+ \psi\alpha_{\Pi n} n_\mu \nabla^{\mu}\alpha  \Big] 
\!-\! \beta n^\mu \Big[ T\nabla_\mu \alpha - \beta_1\dot n_\mu 
- \beta_{nn} n_\mu \theta  \nonumber \\
&+ \alpha_0\nabla_{\mu}\Pi
+ \alpha_1\nabla_\nu \pi^\nu_\mu 
+ \tilde \psi\alpha_{n\Pi } \Pi\dot u_\mu
+ \tilde \psi\alpha_{\Pi n}\Pi\nabla_{\mu}\alpha \nonumber \\ 
&+ \tilde \chi\alpha_{\pi n} \pi^\nu_\mu \nabla_\nu \alpha  
+ \tilde \chi\alpha_{n\pi } \pi^\nu_\mu \dot u_\nu \Big] 
\!\!+\! \beta\pi^{\mu\nu}\Big[ \sigma_{\mu\nu} 
\!-\! \beta_2\dot\pi_{\mu\nu} \nonumber \\
&- \beta_{\pi\pi}\theta\pi_{\mu\nu}
- \alpha_1 \nabla_{\langle\mu}n_{\nu\rangle} 
- \chi\alpha_{\pi n}n_{\langle\mu}\nabla_{\nu\rangle}\alpha \nonumber \\
&- \chi\alpha_{n\pi }n_{\langle\mu}\dot u_{\nu\rangle} \Big],
\end{align}
where $\alpha_i,~\beta_i,~\alpha_{XY},~\beta_{XX}$ are known functions
of $\beta,~\alpha$ and the integral coefficients $I_{nq}^\pm,~J_{nq}^\pm$ and
$K_{nq}^\pm$. Two new parameters $\psi$ and $\chi$ with $\tilde \psi =
1-\psi$ and $\tilde \chi = 1-\chi$ are introduced to `share' the
contributions stemming from the cross terms of $\Pi$ and
$\pi^{\mu\nu}$ with $n^{\mu}$.

The second law of thermodynamics, $\partial_{\mu}S^{\mu}\ge 0$, is
guaranteed to be satisfied if we impose linear relationships between
thermodynamical fluxes and extended thermodynamic forces, leading to
the following evolution equations for bulk, charge current and shear
\begin{align}
\Pi &= -\zeta\Big[ \theta 
+ \beta_0 \dot \Pi 
+ \beta_{\Pi\Pi} \Pi \theta 
+ \alpha_0 \nabla_\mu n^\mu  \nonumber \\
&\quad\quad\quad + \psi\alpha_{n\Pi} n_\mu \dot u^\mu
+ \psi\alpha_{\Pi n} n_\mu \nabla^\mu \alpha  \Big], \label{bulk} \\ 
n^{\mu} &= \lambda \Big[ T \nabla^\mu \alpha 
- \beta_1\dot n^{\langle\mu\rangle} 
- \beta_{nn} n^\mu \theta
+ \alpha_0 \nabla^\mu \Pi \nonumber \\
&\quad\quad\; + \alpha_1 \Delta^\mu_\rho \nabla_\nu \pi^{\rho\nu}
+ \tilde \psi\alpha_{n\Pi} \Pi \dot u^{\langle\mu\rangle}
+ \tilde \psi\alpha_{\Pi n} \Pi \nabla^\mu \alpha \nonumber \\
&\quad\quad\; + \tilde \chi\alpha_{\pi n} \pi_\nu^\mu \nabla^\nu \alpha
+ \tilde \chi\alpha_{n\pi} \pi_\nu^\mu \dot u^\nu \Big], \label{current} \\ 
\pi^{\mu\nu} &= 2\eta\Big[ \sigma^{\mu\nu} 
- \beta_2\dot\pi^{\langle\mu\nu\rangle} 
- \beta_{\pi\pi}\theta\pi^{\mu\nu}
- \alpha_1 \nabla^{\langle\mu}n^{\nu\rangle} \nonumber \\
&\quad\quad\quad - \chi\alpha_{\pi n} n^{\langle\mu} \nabla^{\nu\rangle} \alpha 
- \chi\alpha_{n\pi } n^{\langle\mu} \dot u^{\nu\rangle} \Big] , \label{shear}
\end{align}
with the coefficients of charge conductivity, bulk and shear
viscosity, viz. $\lambda, \zeta,\eta \ge 0$. The notations,
$A^{\langle\mu\rangle} = \Delta^{\mu}_{\nu}A^{\nu}$ and
$B^{\langle\mu\nu\rangle} =
\Delta^{\mu\nu}_{\alpha\beta}B^{\alpha\beta}$ represent space-like and
traceless symmetric projections respectively, both orthogonal to $u^{\mu}$,
where $\Delta^{\mu\nu}_{\alpha\beta} =
[\Delta^{\mu}_{\alpha}\Delta^{\nu}_{\beta} +
  \Delta^{\mu}_{\beta}\Delta^{\nu}_{\alpha} -
  (2/3)\Delta^{\mu\nu}\Delta_{\alpha\beta}]/2$. It may be noted that
although the forms of the Eqs.  (\ref{bulk})-(\ref{shear}) are the
same as in the standard Israel-Stewart theory
\cite{Israel:1979wp,Muronga:2003ta}, all the transport coefficients
are explicitly determined in the present derivation:
\begin{align}
\beta_0 &= \lambda^2_\Pi J_{10}^+ /\beta,\quad 
\beta_1 = -\lambda^2_n J_{31}^+ /\beta,\quad 
\beta_2 = 2\lambda^2_\pi J_{52}^+ /\beta, \nonumber \\
\alpha_0 &= \lambda_\Pi \lambda_n J_{21}^+ /\beta,\quad 
\alpha_1 = -2\lambda_\pi \lambda_n J_{42}^+ /\beta. \label{alphas}
\end{align}
As a consequence, the relaxation times defined as,
\begin{equation}\label{RT}
\tau_{\Pi} = \zeta\,\beta_0, \quad
\tau_{n} = \lambda\,\beta_1, \quad
\tau_{\pi} = 2\,\eta\,\beta_2 ,
\end{equation}
can be obtained directly.
With $\lambda_\Pi = -1/J_{21}^+$, $\lambda_n=1/J_{21}^-$, 
$\lambda_{\pi}=1/(2J_{42}^+)$, $n=I_{10}^-$, $\epsilon=I_{20}^+$, and
$P=-I_{21}^+$, the expressions for $\beta_1,\alpha_0,\alpha_1$ simplify to
\begin{equation}\label{B1A0A1}
\beta_1 = (\epsilon+P)/n^2, \quad   \alpha_0 = \alpha_1 = 1/n. 
\end{equation}
For a classical Boltzmann gas ($\tilde f_0=1$), 
the coefficients $\beta_0$ and $\beta_2$ take the simple forms
\begin{equation}\label{B0B2}
\beta_0 = 1/P,\quad  
\beta_2 = 3/(\epsilon+P) + m^2\beta^2P/[2(\epsilon+P)^2].
\end{equation}
Equations (\ref{bulk})-(\ref{shear}) in conjunction with the
second-order transport coefficients (\ref{B1A0A1}) and (\ref{B0B2})
constitute one of the main results in the present work. These
coefficients are obtained consistently within the same theoretical
framework. In contrast, in the standard derivation from entropy
principles \cite{Israel:1979wp}, the transport coefficients have to be
estimated from an alternate theory. For instance, in the IS derivation
based on kinetic theory, these involve complicated expressions which
in the photon limit ($m \beta \to 0$) reduce to \cite{Israel:1976tn}
\begin{equation}\label{IST}
\beta_0^{IS} = 216/(m^4 \beta^4 P), \quad \beta_2^{IS} = 3/4P.
\end{equation}
An alternate derivation from kinetic theory (KT) using directly the 
definition of dissipative currents yields 
\cite{Denicol:2010xn}
\begin{align}\label{DKR}
\beta_0^{KT} =& \Big[ \left(\frac{1}{3}-c_s^2 \right)(\epsilon+P)-\frac{2}{9}(\epsilon-3P) \nonumber \\
&~ - \frac{m^4}{9}\mean{(u.p)^{-2}} \Big]^{-1},  \nonumber \\
\beta_2^{KT} =& \, \frac{1}{2} \left[ \frac{4P}{5}
+\frac{1}{15}(\epsilon-3P)
- \frac{m^4}{15}\mean{(u.p)^{-2}}\right]^{-1},
\end{align}
where $c_s$ is the speed of sound and $\mean{\cdots}\equiv \int dp(\cdots)f_0$.
A field-theoretical (FT) approach gives \cite{Huang:2011ez}
\begin{align}\label{HK}
\beta_0^{FT} =& \left[ \left(\frac{1}{3}-c_s^2 \right)(\epsilon+P)-\frac{a}{9}(\epsilon-3P)\right]^{-1},
\nonumber \\
\beta_2^{FT} =& \, 1/[2(3-a)P],
\end{align}
where $a=2$ for charged scalar bosons and $a=3$ for fermions.  We find
that our expression for $\beta_2$ (Eq. (\ref{B0B2})) in the massless
limit, agrees with the IS result (Eq. (\ref{IST})) and also with those
obtained in Refs. \cite{Baier:2006um,El:2009vj}. Thus the shear
relaxation times $\tau_\pi$ (Eq. (\ref{RT})) obtained here and in
these studies are also identical. As $\beta_0$ in
Eqs. (\ref{IST})-(\ref{HK}) diverge in the massless limit, so does
the bulk relaxation time $\tau_\Pi$ (Eq. (\ref{RT})), thereby stopping
the evolution of the bulk pressure. It is important to note that
$\beta_0$ in Eq. (\ref{B0B2}) and hence $\tau_\Pi$ in the present
calculation remain finite in this limit. A detailed comparison of IS, KT
and FT results can be found in \cite{Denicol:2010br}. The two
parameters $\psi$ and $\chi$ occurring in Eq. (\ref{EFCD3}) remain
undetermined as in \cite{Israel:1979wp}; however, these do not
contribute to the scaling expansion.

\begin{figure}[t]
\begin{center}
\scalebox{0.35}{\includegraphics{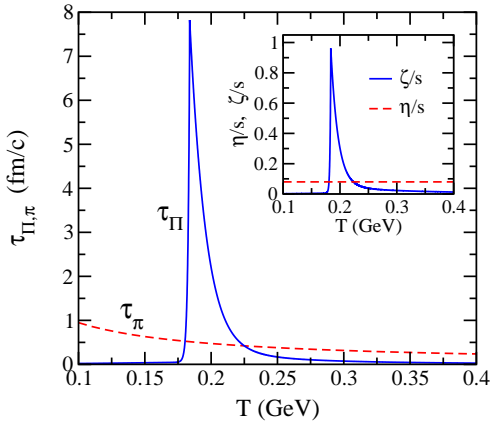}}
\end{center}
\vspace{-0.4cm}
\caption{(Color online) Temperature dependence of bulk and shear relaxation times. 
Inset shows $\zeta/s$ (see text) and $\eta/s = 1/4\pi$.}
\label{tauT}
\end{figure}

To demonstrate the numerical significance of the new coefficients
derived here, we consider the evolution equations in the
boost-invariant Bjorken hydrodynamics at vanishing net baryon number
density \cite{Bjorken:1982qr}. In terms of the coordinates
($\tau,x,y,\eta$) where $\tau = \sqrt{t^2-z^2}$ and
$\eta=\tanh^{-1}(z/t)$, the initial four-velocity becomes
$u^\mu=(1,0,0,0)$. For this scenario $n^\mu=0$ and the evolution
equations for $\epsilon$, $\pi \equiv -\tau^2 \pi^{\eta \eta}$ and
$\Pi$ reduce to
\begin{align}
\frac{d\epsilon}{d\tau} &= -\frac{1}{\tau}\left(\epsilon + P + \Pi -\pi\right), \label{BED} \\
\tau_{\pi}\frac{d\pi}{d\tau} &= \frac{4\eta}{3\tau} - \pi - \frac{4\tau_{\pi}}{3\tau}\pi, \label{Bshear} \\
\tau_{\Pi}\frac{d\Pi}{d\tau} &= -\frac{\zeta}{\tau} - \Pi - \frac{4\tau_{\Pi}}{3\tau}\Pi \label{Bbulk}.
\end{align}
Noting that $\beta_0=1/P$, $\beta_2=3/(\epsilon+P)$ and $s=(\epsilon+P)/T$,
the relaxation times defined in Eq. (\ref{RT}) reduce to
\begin{equation}\label{Mtaus} 
\tau_{\Pi} = \frac{\epsilon+P}{PT}\left(\frac{\zeta}{s}\right), \quad
\tau_{\pi} = \frac{6}{T}\left(\frac{\eta}{s}\right).
\end{equation}

We have used the state-of-the-art equation of state
\cite{Huovinen:2009yb}, which is based on a recent lattice QCD result
\cite{Bazavov:2009zn}. For $\zeta/s$ at $T \geq T_c \approx 184$ MeV,
we use the parametrized form \cite{Rajagopal:2009yw} of the lattice
QCD results of Meyer \cite{Meyer:2007dy} which suggest a peak near
$T_c$. At $T<T_c$, the sharp drop in $\zeta/s$ reflects its extremely
small value found in the hadron resonance gas model \cite{Prakash:1993bt};
see inset of Fig. \ref{tauT}. For the $\eta/s$ ratio, we use the
minimal KSS bound \cite {Kovtun:2004de} value of $1/4\pi$.

In the absence of any reliable prediction for the bulk relaxation time
$\tau_\Pi$, it has been customary to keep it fixed or set it equal to
the shear relaxation time $\tau_\pi$ or parametrize it in such a way
that it captures critical slowing-down of the medium near $T_c$ due to
growing correlation lengths
\cite{Fries:2008ts,Denicol:2009am,Song:2009rh,Rajagopal:2009yw}.
Since $\zeta/s$ has a peak near the phase transition, the $\tau_\Pi$
obtained here (Eq. (\ref{Mtaus})) and shown in Fig. \ref{tauT}, 
{\it naturally} captures the phenomenon of critical slowing-down.

The evolution equations (\ref{BED})-(\ref {Bbulk}) are solved
simultaneously with an initial temperature $T_0 = 310$ MeV
\cite{Rajagopal:2009yw} and initial time $\tau_0 = 0.5$ fm/c typical
for the RHIC energy scan. We take initial values for bulk stress and
shear stress, $\Pi= \pi = 0$ GeV/fm$^3$ which corresponds to an
isotropic initial pressure configuration.

\begin{figure}[t]
\begin{center}
\scalebox{0.37}{\includegraphics{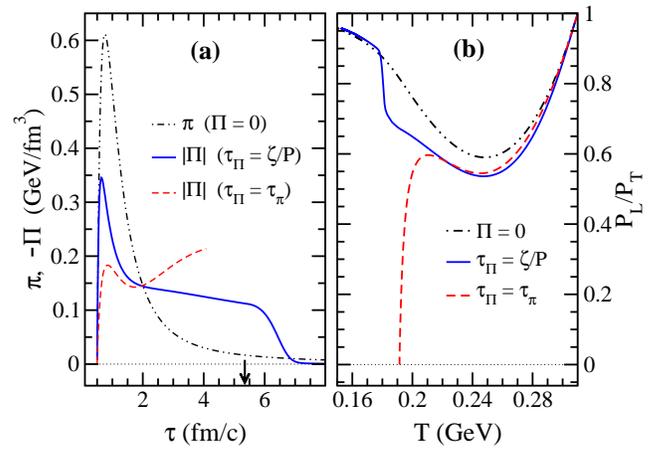}}
\end{center}
\vspace{-0.4cm}
\caption{(Color online) (a) Time evolution of shear stress in the absence of bulk
  ($\Pi=0$) and magnitude of bulk stress for $\tau_\Pi=\zeta/P$ and
  $\tau_\Pi = \tau_\pi$. The arrow indicates the time when $T_c$ is
  reached. (b) Temperature dependence of pressure anisotropy,
  $P_L/P_T$, for these three cases. The results are for initial
  $T=310$ MeV, $\tau_0=0.5$ fm/c and $\eta/s = 1/4\pi$. The evolution
  is stopped when $P_L$ vanishes.}
\label{Piphi}
\end{figure}

Figure \ref {Piphi}(a) shows time evolution of the shear pressure
$\pi$ and the magnitude of the bulk pressure $\Pi$. At early times
$\tau \lesssim 2$ fm/c or equivalently at $T \gtrsim 1.2 T_c$, shear
dominates bulk. This implies that eccentricity-driven elliptic flow
which develops early in the system would be controlled more by the
shear pressure \cite{Song:2009rh}. At later times (when $T \sim T_c$),
the large value of $\zeta/s$ makes the bulk pressure dominant. This
leads to sizeable entropy generation (Eq. (\ref{EFCD3})) and
consequently enhanced particle production.

Figure \ref {Piphi}(a) also compares the $\Pi$ evolution for bulk
relaxation time, $\tau_{\Pi}$, calculated from Eq. (\ref{Mtaus})
(solid line) and $\tau_{\Pi} = \tau_{\pi}$ (dashed line). At early
times, the larger value of $\tau_{\Pi}$ in the latter case (see
Fig. \ref{tauT}) results in a relatively smaller growth of $|\Pi|$ as
evident from Eq. (\ref{Bbulk}). Near $T_c$, the rapid increase in
$\zeta/s$ causes $|\Pi|$ to increase. Subsequently the longitudinal
pressure $P_L= (P+\Pi-\pi)$ vanishes leading to cavitation
\cite{Fries:2008ts,Rajagopal:2009yw,Torrieri:2008ip,Bhatt:2010cy}. In
contrast, with our $\tau_{\Pi}$, this rise in $\zeta/s$ is
overcompensated by a faster increase in $\tau_{\Pi}$ thereby slowing
down the evolution of $\Pi$. This behavior prevents the onset of
cavitation and guarantees the applicability of hydrodynamics with bulk
and shear up to temperatures well below $T_c$ into the hadronic
phase. Furthermore, this slowing down of the medium followed by its
rapid expansion, has the right trend to explain the identical-pion
correlation measurements (Hanbury Brown-Twiss puzzle)
\cite{Paech:2006st,Pratt:2008qv}.

The absence of cavitation in our calculation is clearly evident in
Fig. \ref{Piphi}(b) which shows the variation of pressure anisotropy,
$P_L/P_T = (P+\Pi-\pi)/(P+\Pi+\pi/2)$, with temperature. Near $T_c$,
the longitudinal pressure $P_L$ vanishes if one assumes $\tau_{\Pi} =
\tau_{\pi}$ (dashed line) leading to cavitation, whereas it is found
to be positive for all temperatures with $\tau_\Pi$ derived here
(solid line). In fact, we have found that in the latter case,
cavitation is completely avoided for the entire range of $\zeta/s$
values ($0.5 < \zeta/s < 2.0$ near $T_c$) estimated in lattice QCD
\cite{Meyer:2007dy}. The sizeable difference between the $\Pi
= 0$ case (dot-dashed line) and the $\tau_\Pi=\zeta/P$ case (solid
line) clearly underscores the importance of bulk pressure near $T_c$,
which can have significant implications for the elliptic flow $v_2$
\cite{Denicol:2009am} thus affecting the extraction of $\eta/s$.
Further, the large bulk pressure when incorporated in the freezeout
prescription could also affect the final particle abundances and
spectra.

We have also found that the evolution of $\Pi$ is insensitive to the
choice of initial conditions such as $\Pi(\tau_0)=0$ and the
Navier-Stokes value $-\zeta(T_0)/\tau_0$. This is due to very small
$\tau_\Pi$ at early times (or higher temperatures) which causes $\Pi$
to quickly lose the memory of its initial condition and to relax to
the same value at $\tau \gtrsim 1$ fm/c.

To summarize, we have presented a new derivation of the relativistic
dissipative hydrodynamic equations from entropy considerations. We
arrive at the same form of dissipative evolution equations as in the
standard derivation but with all second-order transport coefficients
such as the relaxation times and the entropy flux coefficients
determined consistently within the same framework. We find that in the
Bjorken scenario, although the bulk pressure can be large, the
relaxation time derived here prevents the onset of cavitation due to
the critical slowing down of bulk evolution near $T_c$.

\end{document}